\documentclass[12pt]{article}
\usepackage{amsmath}
\usepackage{amsfonts}
\usepackage{amssymb}
\usepackage{mltex}
\usepackage{graphicx}
\newtheorem{lemma}{Lemma}[section]
\newtheorem{theorem}[lemma]{Theorem}
\newtheorem{proposition}[lemma]{Proposition}

\newtheorem{remark}[lemma]{Remark}

\newtheorem{definition}[lemma]{Definition}

\def\sq{\hbox {\rlap{$\sqcap$}$\sqcup$}}
\def\sq{\hbox {\rlap{$\sqcap$}$\sqcup$}}

\def\1{{\rm 1\mskip-4.5mu l} }
\def\lsim{\raise0.3ex\hbox{$<$\kern-0.75em\raise-1.1ex\hbox{$\sim$}}}
\def\gsim{\raise0.3ex\hbox{$>$\kern-0.75em\raise-1.1ex\hbox{$\sim$}}}

\def\beq{\begin{equation}}   \def\edq{\end{equation}}
\def\bea{\begin{eqnarray}}  \def\eea{\end{eqnarray}}

\newcommand\mysection{\setcounter{equation}{0}\section}
\renewcommand{\theequation}{\thesection.\arabic{equation}}
\newcounter{hran} \renewcommand{\thehran}{\thesection.\arabic{hran}}

\def\bmini{\setcounter{hran}{\value{equation}}
    \refstepcounter{hran}\setcounter{equation}{0}
    \renewcommand{\theequation}{\thehran\alph{equation}}\begin{eqnarray}}

\def\bminiG#1{\setcounter{hran}{\value{equation}}
\refstepcounter{hran}\setcounter{equation}{-1}
\renewcommand{\theequation}{\thehran\alph{equation}}
\refstepcounter{equation}\label{#1}\begin{eqnarray}}

\def\emini{\end{eqnarray}\relax\setcounter{equation}{\value{hran}}\renewcommand{
\theequation}
{\thesection.\arabic{equation}}}

\begin{document}

\title { Quantum and Classical Fidelity for Singular Perturbations of the Inverted and 
Harmonic Oscillator}

\author{\it{\bf Monique Combescure}\\
\it IPNL, B\^atiment Paul Dirac \\
\it 4 rue Enrico Fermi,Universit\'e Lyon-1 \\
\it  F.69622 VILLEURBANNE Cedex, France\\
\tt email monique.combescure@ipnl.in2p3.fr\\
\it and\\
\it {\bf Alain Combescure} \\
\it LaMCoS, INSA-LYON\\
\it B‰timent Jean d'Alembert\\
\it18-20 rue des Sciences\\
\it F-69621 VILLEURBANNE Cedex, France}
\vskip 1 truecm
\maketitle

\bigskip

\begin{abstract}

Let us consider the quantum/versus classical dynamics for Hamiltonians of the form
\beq
\label{0.1}
H_{g}^{\epsilon} := \frac{P^2}{2}+ \epsilon \frac{Q^2}{2}+ \frac{g^2}{Q^2}
\edq
where $\epsilon = \pm 1$, $g$ is a real constant. We shall in particular study the Quantum
 Fidelity between $H_{g}^{\epsilon}$ and $H_{0}^{\epsilon}$ defined as 
 \beq
 \label{0.2}
 F_{Q}^{\epsilon}(t,g):= \langle \exp(-it H_{0}^{\epsilon})\psi, exp(-itH_{g}^
 {\epsilon})\psi
 \rangle
 \edq
for some reference state $\psi$ in the domain of the relevant operators.\\

We shall also propose a definition of the Classical Fidelity, already present in the literature 
(\cite{becave1}, \cite{becave2}, \cite{ec}, \cite{prozni}, \cite{vepro}) and compare
 it with the behaviour of the Quantum Fidelity, as time evolves, and
as the coupling constant $g$ is varied.
\end{abstract}
\section{INTRODUCTION}

In the last few years, there has been a renewal of interest in the notion of ``Quantum 
Fidelity'' (also called Loschmidt Echo), in particular for applications in Quantum Chaos or
Quantum Computation problems, (see for example ref. \cite{per}, \cite{becave3}, 
\cite{goprose} and
 \cite{cu} for 
 recent reviews
including many earlier references ).
This notion is very simple because it amounts to considering the behaviour in time of the overlap
of two quantum states: one evolved according to a given dynamics, the other one evolved by
a slight perturbation of it, but starting from the {\bf same initial state } at time zero. While
this overlap obviously equals one at time zero, it starts decreasing as time evolves, although
rather slightly if the size of the perturbation is small.
Thus this phenomenon should rather be described as ``sensitivity to perturbations'' than
by the naive denomination ``fidelity'', but we use it nevertheless since it has now become very
common in the literature of the subject.\\

 It is believed (and sometimes shown
numerically) that the ``generic'' limiting value is zero, and that the type of decay (Gaussian
or exponential) strongly depends on the chaotic versus regular classical underlying dynamics.
However for some integrable systems strong recurrences to exact fidelity have been shown
\cite{sala}. All these approaches are however rather heuristic, and these questions have not
been  treated rigorously, except recently by one of the authors \cite {co1}.\\

In this paper we pursue a rigorous study of the Quantum Fidelity problem for rather simplistic
Hamiltonian systems : Harmonic and Inverse Harmonic Oscillator perturbed by repulsive
 inverse quadratic potentials. The reference quantum states considered belong to a rather large
 class namely Perelomov's Generalized Coherent states of the $SU(1,1)$ algebra, suitable for
  the dynamics
 under consideration. This class, however large, is very specific, in particular the quantum
 (perturbed as well as unperturbed) dynamics can be exactly solved in terms of the classical
 trajectories of the {\bf unperturbed} classical dynamics. Then the quantum fidelity can be
 shown to equal in absolute value the ``return probability'' for a very simple quantum
 dynamics and elementary reference states. Two very different behaviours in time are then
 demonstrated:\\
 $\bullet$ periodic recurrences to 1 of the quantum fidelity for Harmonic Oscillator
 unperturbed dynamics\\
 $\bullet$ exponential decrease in time to some asymptotic (non-zero) value in the future as
 well in the past for the Inverted Harmonic Oscillator unperturbed dynamics.\\
 
 Then a similar notion already present in the physical literature (\cite{becave1}, 
 \cite{becave2},
 \cite{ec}, \cite{prozni}, \cite{vepro}) of ``Classical Fidelity'' is considered for these
  systems. It is just the overlap of two distribution functions in classical phase-space, one
  evolved by the unperturbed, the other one by the unperturbed classical dynamics. For the
  particular case of dynamics considered in this paper, we can evaluate the long time behaviour
  of the Classical Fidelities for different distribution functions.\\
  It is shown that a similar behaviour as for the quantum fidelities occurs, namely:\\
  $\bullet$ periodic recurrences to 1 in the H. O. case\\
  $\bullet$ fast decrease to some non-zero asymptotic value in the future as well in the past
  in the I. H. O. case.\\
  However some differences, in particular in the neighborhood of initial time are demonstrated.
\\

The plan of the paper is as follows:\\
In Section 2 we show that the quantum dynamics is exactly solvable in terms of the classical
one for both the H. O. and I. H. O., and we introduce the class of quantum reference states
under consideration. In section 3 (resp. 4),we describe the Quantum (resp. Classical) Fidelities
for the H. O. case (stable case). In section 5 (resp. 6), we describe the Quantum (resp. Classical)
Fidelities for the I. H. O. case (unstable case). In Section 7 we give concluding remarks.
The Appendix contains the Proofs of the various estimates provided in Sections 4 and 6.\\

\mysection{Quantum Fidelity for a suitable class of reference states (Perelomov
 Generalized Coherent States)}

In all this section $\epsilon= \pm 1$. Furthermore the Hamiltonian $H_{g}^{+1}$ (resp.
$H_{g}^{-1}$ ) of equation (0.1) is simply denoted  $H_{g}^+$ (resp. $H_{g}^-$)\\
 
According to \cite{pe}, the evolution operator $\exp(-itH_{g}^{\epsilon})\psi$ can be 
expressed explicitly
in terms of solutions of the classical motion for $H_{0}^{\epsilon }$, for any $g$. 
Here $\psi$ 
belongs to a suitable class of wavepackets that we shall make precise below. 
\\
\noindent
Let us denote by $z_{\epsilon}(t)$ the general form of a complex solution for Hamiltonian $H_{0}^
{\epsilon}$:
\beq
\label{1.0}\ddot z_{\epsilon}(t)+\epsilon z_{\epsilon}(t)=0
\edq
 Its polar decomposition is written as:
\beq
\label{1.1}
z_{\epsilon}(t):= \exp (u_{\epsilon}+i \theta_{\epsilon})
\edq
with $t \mapsto u_{\epsilon}, \ \theta_{\epsilon}$ being {\bf real functions}. 
The constant wronskian of $z_{\epsilon}$ and $\bar z_{\epsilon}$ is taken as $2i $. 
This yields:
 \beq
 \label{1.2}
 \dot \theta_{\epsilon}= e^{-2u_{\epsilon}}
 \edq
From equation (\ref{1.0}), we easily deduce that $u_{\epsilon}(t)$ obeys the following 
differential equation:
\beq
\label{1.3}
\ddot u_{\epsilon}+ \dot u_{\epsilon}^2-e^{-4u_{\epsilon}}+\epsilon=0
\edq
Let us denote by $D(u,v)$ the following unitary operator:
\beq
\label{1.4}
D(u,v):= \exp\left(\frac{-iv Q^2}{2}\right)\exp\left(\frac{iu(Q.P+P.Q)}{2}\right)
\edq
We shall now choose as reference wavepackets $\psi$ the Generalized Coherent States for the
$SU(1,1)$ algebra (see \cite{pe}). This means that we consider the state $\psi_{0}$ which is
the ground state of the operator $H_{g}^{+}$
namely
\beq
\label{1.6}
\psi_{0}(x):= c_{g} x^{\alpha}e^{-x^2/2}
\edq
$c_{g}$ being a normalization constant such that $\Vert \psi_{0}\Vert =1$ and $\alpha$ being
determined by
\beq
\label{1.7}
\alpha := \frac{1}{2}+ \sqrt {\frac{1}{4}+2g^2}
\edq
one has
\beq
\label{1.8}
H_{g}^{+}\psi_{0}=( \alpha + \frac{1}{2})\psi_{0}
\edq
We know from \cite{pe} that $\psi$ has the following general form
\beq
\label{1.9}
\psi_{u_{0}, v_{0}, \theta_{0}}
 := \exp \left(-i(\alpha+\frac{1}{2})\theta_{0}\right)D(u_{0},v_{0})\psi_{0}
\edq
for general {\bf real constants} $\theta_{0},\ u_{0},\ v_{0}$.\\

Then we have proven the general result (see \cite{co1}):

\begin{proposition}
Let $u_{\epsilon}, \ \theta_{\epsilon}$ be the functions defined above. 
Assume that they have the initial data
\beq
\label{1.10}
\theta_{\epsilon}(0)=\theta_{0},\ u_{\epsilon}(0)= u_{0},\ 
\dot u_{\epsilon}(0)= v_{0}
\edq
Then for any $g$ we have:
\beq
\label{1.11}
e^{-it H_{g}^{\epsilon}}\psi = \psi_{u_{\epsilon}(t), \dot u_{\epsilon}(t),
 \theta_{\epsilon}(t)} \equiv
\exp\left(-i\theta_{\epsilon}(t)(\alpha+
\frac{1}{2})\right)
D(u_{\epsilon}(t), \dot u_{\epsilon}(t))\psi_{0}
\edq
which means that the set of Generalized Coherent States is stable under the quantum evolution
generated by $H_{g}^{\epsilon}$.
Moreover
\beq
\label{1.12}
e^{-it H_{0}^{\epsilon}}\psi = e^{-i\theta_{0}(\alpha + \frac{1}{2})}
D(u_{\epsilon}(t), \dot
u_{\epsilon}(t))\exp(-i(\theta_{\epsilon}(t)-\theta_{0})H_{0}^+)\psi_{0}
\edq
\end{proposition}

\begin{remark}
The  {\bf same} functions $u_{\epsilon}, \ v_{\epsilon}(t)$ and
 $\theta_{\epsilon}$ appear in the
formulas (\ref{1.11}) and (\ref{1.12}) above.
\end{remark}

We have the following important result:

\begin{theorem}
For any real $g$ and for $\epsilon = \pm1$, we have:
\beq
\label{1.15}
 F_{Q}^{\epsilon}(t,g)= e^{-i \tilde\theta_{\epsilon}(t)(\alpha+ 1/2)}
 \langle \psi_{0}, 
 \exp (i\tilde\theta_{\epsilon}(t)H_{0}^+)\psi_{0}\rangle
 \edq
 where by $\tilde \theta_{\epsilon}(t)$ we have denoted:
 \beq
 \label{1.16}
 \tilde\theta_{\epsilon}(t):= \theta_{\epsilon}(t)- \theta_{0}
 \edq
\end{theorem}

Proof: This follows easily from equation (\ref {1.11}), and (\ref{1.12}), and using the 
unitarity of the operator $D(u,v)$.\\
\sq\\

Thus the important fact to notice is that the modulus of the Quantum Fidelity (which is just
the quantity refered to as Quantum Fidelity in the literature) reduces to the so-called 
{\bf ``return probability''} in the state $\psi_{0}$ for the reference states
 $\psi_{u_{0}, v_{0}, \theta_{0}}$ under 
consideration in this paper, for some rescaled time $\tilde\theta_{\epsilon}(t)$.\\

From now on, the modulus of the Quantum Fidelity functions for the case $\epsilon = +1$
 (resp. $\epsilon =-1$) will be denoted 
as $F_{Q}(t,g)$ (resp. $G_{Q}(t,g)$). Thus:
\beq
\label{1.14}
F_{Q}(t,g)= \vert F_{Q}^{+1}(t,g)\vert,\quad G_{Q}(t,g)=\vert F_{Q}^{-1}(t,g)
\vert
\edq

\mysection{Behavior of the Quantum Fidelity for $\epsilon =+1$}

$H_{0}^{+}= \frac{P^2+Q^2}{2}$ is simply the Harmonic Oscillator, whose classical 
solutions
are linear combinations of $\cos t$ and $\sin t$.\\

The most general form of a complex solution is:
\beq
\label{2.1}
z(t):= (a +ib)\cos t + (c+id)\sin t
\edq
with $a,\ b,\ c,\ d \in \mathbb R$.
The constant wronskian of $z$ and $\bar z$ is taken as $2i $. This yields:
 \beq
 \label{2.2}
 ad-bc = 1
 \edq
Writing $z_{+}:= e^{u_{ +}+i\theta_{ +}}$, as in the previous section we get:

\beq
\label{2.3}
u_{ +}(t):= \frac{1}{2}\log \{(a\cos t + c \sin t)^2 +(b \cos t + d \sin t)^2\}
\edq
\beq
\label{2.4}
\tan \theta_{ +}(t) := \frac{b \cos t + d \sin t}{a \cos t + c \sin t}
\edq
Thus:
\beq
\label{2.5}
u_{ +}(0)= \frac{1}{2}\log (a^2+b^2), \quad \dot u_{+}(0)= \frac{ac+bd}{a^2+b^2},
\quad \theta_{+}(0)= \arctan(\frac{b}{a})
\edq

Clearly we have the following general result for $F_{Q}^{+1}(t,g)$:

\begin{proposition}
\beq
\label{2.8}
F_{Q}^{+1}(t,g)= e^{-i \alpha \tilde\theta_{+}(t)}\sum_{n=0}^{\infty}\vert 
\lambda_{n}
\vert ^2 e^{in \tilde\theta_{+}(t)}
\edq
where $\lambda_{n}$ are the coefficients of the expansion of $\psi_{0}$ in the eigenstates
$\phi_{n}$ of $H_{0}^{+}$, and of course
\beq
\sum_{n=0}^{\infty} \vert \lambda_{n}\vert ^2=1
\edq
This expansion is finite and involves only even terms $\exp(2in\tilde\theta_{+}(t))$ 
in the 
particular case where $g$ is of the form $g= \sqrt {k(k+1)/2}$ for $k=1,2, ...$.
\end{proposition}
The proof is an immediate consequence of equation (\ref{1.15}).\sq\\

Recall that 
\beq
\phi_{n}(x)= (\sqrt \pi n! 2^n)^{-1/2}e^{-x^2/2}H_{n}(x)
\edq
where $H_{n}$ are the Hermite polynomials
$H_{n}(x)= (-1)^n e^{x^2}\left(\frac{d}{dx}\right)^n e^{-x^2}$
\\

Let us consider the following cases $g=1, \sqrt 3 , \sqrt {10}$ (which yields $\alpha = 1, 2
, 3$ respectively). Then we have
\beq
F_{Q}^{+1}(t,1)= \frac{2}{3}+ \frac{1}{3}\exp(2i\tilde\theta_{+}(t)) \qquad
F_{Q}^{+1}(t, \sqrt 3) = \frac{2}{5}+ \frac{3}{5}\exp(2i \tilde\theta_{+}(t))
\edq
\beq
F_{Q}^{+1}(t, \sqrt {10})= \frac{8}{35}+ \frac{24}{35} e^{2i\tilde\theta_{+}(t)}+
\frac{3}{35} e^{4i\tilde\theta_{+}(t)}
\edq

\bigskip
{\bf Study of $\tilde \theta_{+}(t)$}
\\

Of course $F_{Q}(t,g)$ depends on the reference state $\psi$ via the coefficients
 $\lambda_{n}$.
Clearly  $\tilde\theta_{+}(t)$ is {\bf independent of $g$} and we have the following 
property:

\begin{lemma}
$\tilde \theta_{+} (2\pi)= 2\pi$, and thus $F_{Q}(t,g)$ is $2\pi$-periodic in $t$ for any 
$g \in \mathbb R$. Moreover for $g=1,\ \sqrt 3,\ \sqrt{10}$,
\beq
F_{Q}(k\pi,g)=1, \qquad \forall k \in \mathbb Z
\edq
\end{lemma}

 Proof:
 \beq
 \int_{0}^{\pi}\ ds \ e^{-2u_{+}(s)}=  
 \int_{-\pi/2}^{+\pi/2}\ ds \ e^{-2u_{+}(s)}=  \int_{-\infty}^{+\infty}\frac{dx}{(c^2+d^2)(x+\frac{ac+bd}{c^2+d^2})^2
 + \frac{1}{c^2+d^2}} = \pi
 \edq
 \sq
\\
$\tilde \theta(k\pi) = 0, \ (\mbox{mod}\ \pi)$
which implies that for $g=1, \ \sqrt 3, \ \sqrt 10$..., $F_{Q}(t,g)$ is $\pi$-periodic.\\
The minimum (in absolute value) is attained when $\tilde \theta = \frac{\pi}{2}$, ie 
for those values of $t$
such that
\beq
\tan \theta(t) = \frac{b+d \tan t}{a+c \tan t}=-\frac{1}{ \tan \theta(0)}
\edq
which holds if and only if
$t =- \arctan \left(\frac{a^2+b^2}{ac+bd}\right)= - \arctan\left( \frac{1}{\dot u
(0)}\right)$
We thus have
\beq
 \min F_{Q}(t,1)= \frac{1}{3}, \quad \min F_{Q}(t, \sqrt 3)= \frac{1}{5}, \quad
 \min  F_{Q}(t, \sqrt {10}) = \frac{13}{35}
 \edq
 
 We now present the picture of the modulus of the Quantum Fidelity for $g=1$, and constants
 $a,\ b,\ c,\ d$ chosen as $a=d=-c=1, \ b=0$:

 \bigskip
 \includegraphics{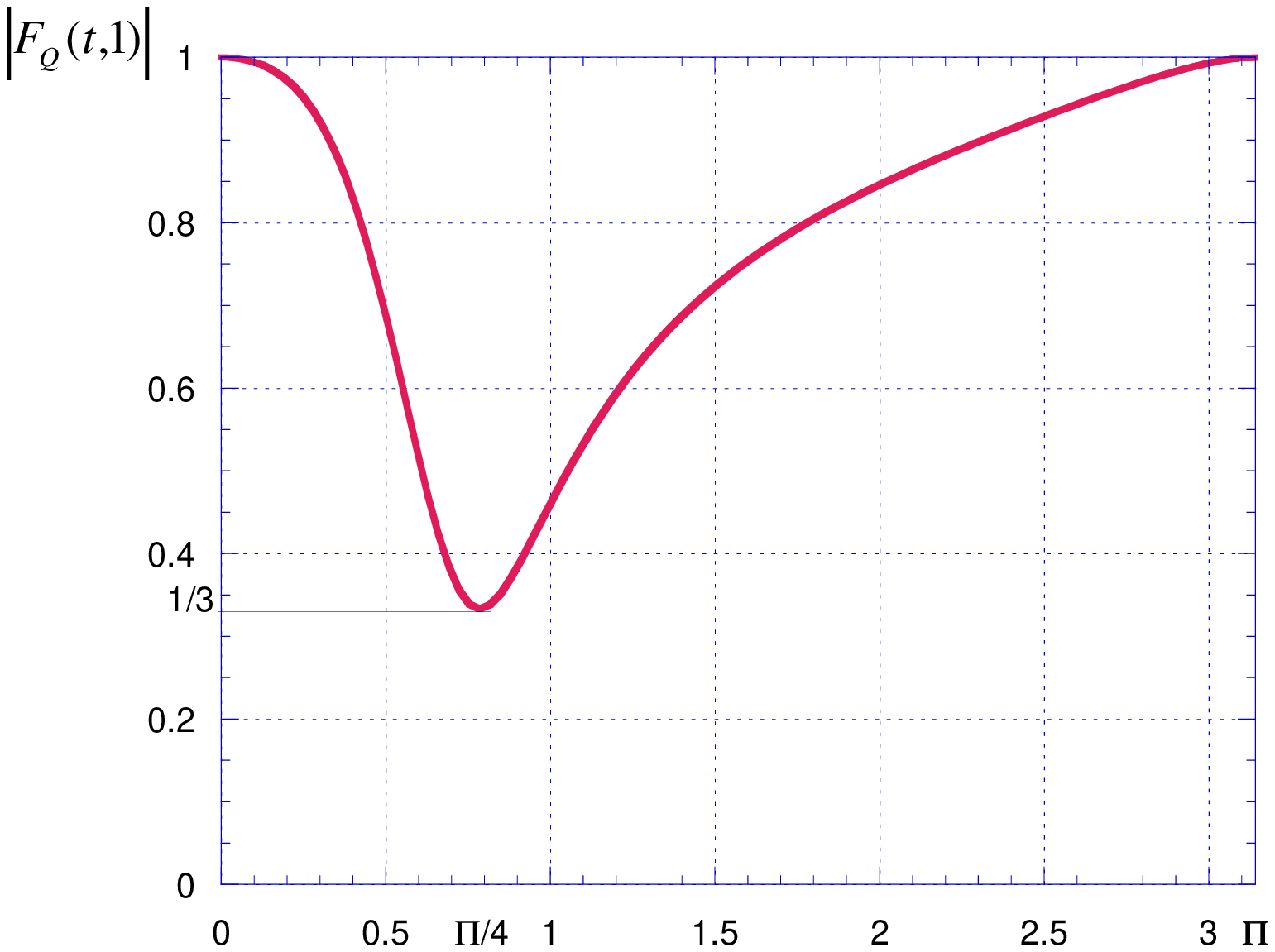}
 \begin{center}
 Fig. 1 : Quantum Fidelity (Stable Case)
 \end{center}
 
 A more symmetrical picture is obtained by taking $a=d=1,\quad b=c=0$, in which case 
 $\tilde\theta_{+}(t)\equiv t$ , and the minimum of $F_{Q}(t,g)$ for $g=1,\ \sqrt 3,
 \ \sqrt{10}....$ is attained for $t= \pi/2$.

 \mysection{Comparison with the Classical Fidelity for $H_{g}^{+}$}
 
 In the literature various definitions of the Classical Fidelity have been proposed and studied,
 (see \cite{becave1}, \cite{becave2}, \cite{ec}, \cite{prozni}, \cite{vepro}).
  Let us here introduce the notion and make a choice of Definition, for which 
 some exact estimates can be performed, together with numerical computations, in order
 to compare its behavior in time with the corresponding Quantum Fidelity.\\
 
 We first show that the classical trajectories for $H_{g}^{+}$ can be simply deduced from
  those
 for $H_{0}^{+}$, and that a natural scaling makes them independent from the constant $g$.
 Let us make the scaling $(q,p) \mapsto (g \sqrt 2)^{1/2}(q,p)$.\\
 
 \begin{proposition}
 Consider  the complex solution (\ref{2.1}) of $\ddot z + z =0$ with wronskian of $z, \bar z$ equal to
 $2i$. Take
 $a=q\ne 0,\ b=0,\ c=p,\ d=\frac{1}{q}$.
  Then we have:
 \beq
 \label{3.1}
 y(t):= \left((q\cos t + p \sin t)^2 + \frac{\sin^2 t}{q^2}\right)^{1/2}
 \edq
 is a solution of differential equation
 $\ddot y + y -\frac{1}{y^3}=0$
 and it has the same initial data as $x(t):=q\cos t + p \sin t$, namely
 $y(0)= x(0), \quad \dot y(0)= \dot x(0)$
 \end{proposition}
 
 Proof:
 An easy computation, using that $\dot \theta(t)= e^{-2u}$ shows that:
 \beq
 \ddot z = \left[\dot u^2 + \ddot u - e^{-4u}+i(\ddot \theta + 2 \dot u \dot \theta)
 \right] = (\dot u^2 + \ddot u - e^{-4u})z
 \edq
 and thus $\ddot z + z=0$ implies $e^{u}(1 + \dot u^2 + \ddot u)- e^{-3u}=0$,
 whence the result, using that $y=e^{u}$ is such that
  $\ddot y = (\ddot u + \dot u^2)y$.
 Furthermore it is easy to check that $y=e^{u}$ is nothing but  (\ref{3.1}).
 \\
 \sq\\
 
 \begin{remark}
 Reintroducing the scaling: $y' = (g\sqrt2)^{1/2}y$, we check that 
$y'(t):= \left((q\cos t + p \sin t)^2 +\frac{2g^2 \sin^2 t}{q^2}\right)^{1/2}$
 is a trajectory for the 
 Hamiltonian $H_{g}^{+}$, namely
 obeys
$ \ddot y' + y' -\frac{2g^2}{y'^3}=0$.
Thus $(y'(t), \dot y'(t))$ is the classical phase space point of classical trajectory for
 $H_{g}^{+}$ that emerges from the same initial point $(q,p)$ as $(q(t), p(t))$, 
 (by continuity we set $y'(0)=0$ if $q=0$). 
 \end{remark}

 We now define the Classical Fidelity as:
 \beq
 F_{C}(t,g):= \int_{\mathbb R^2} \rho(p(t),q(t))\ \rho(\dot y'(t), y'(t)) \ dq dp
 \edq
 where $\rho $ are suitably defined distribution functions in classical phase, satisfying 
 \beq
 \int_{\mathbb R^2}\ dp\ dp \ \rho^2(p,q)=1
 \edq
 
  We shall make
 two different choices:
 \\
 
 1) $\rho(p,q) = G(p,q)\equiv \frac{1}{\sqrt{\pi}} \exp \left( -  \frac{p^2 +q^2}
 {2}\right)$ in which case the Classical Fidelity is denoted $F_{C}(t,g)$
 \\
 
 2) $\rho(p,q) = X(p,q) \equiv \frac{1}{\sqrt{\pi}} \chi(p^2+q^2 \le1)$ in which case 
 the Classical fidelity is denoted $\tilde F_{C}(t, g)$, where $\chi$ is the characteristic 
 function of the set indicated (here a disk of radius 1).\\
 
 \noindent 
 Clearly (due to the parity of $\rho$), $F_{C}(t, g)$ and $\tilde F_{C}(t,g)$ are
  $\pi$- periodic and we have
 \beq
 F_{C}(0, g)=\tilde F_{C}(0,g)=1 ,\  \forall g \in \mathbb R\ {\rm and}\ 
 \tilde F_{C}(t,0) =F_{C}(t,0) \equiv 1, \ \forall t
 \edq 

  Since the Energy is conserved, we have 
$ p(t)^2+q(t)^2 = p^2+q^2 $
 and
 \beq
 y'(t)^2+ \dot y'(t)^2 = p^2+q^2- \frac{2g^2}{y'(t)^2}+ \frac{2g^2}{q^2}
 \edq
 so that:
 \beq
 F_{C}(t) = \frac{1}{\pi} \int_{\mathbb R^2}\ dq \ dp\  \exp\left( -q^2 - p^2 
 + \frac{g^2}{(q\cos t +p \sin t )^2+ \frac{2g^2}{q^2}\sin ^2 t}- 
 \frac{g^2}{q^2}
 \right)
 \edq
 \beq
 \label{3.7}
 \tilde F_{C}(t) = \frac{1}{\pi}\int_{p^2+q^2\le 1}\ dq \ dp \ \chi(y^2(t)+ \dot
 y(t)^2 \le 1)
 \edq
 $F_{C}(t)$ is minimum for $t= \frac{\pi}{2}$  and its minimum equals
 \beq
 \label{3.3}
 F_{C}(\frac{\pi}{2})= \frac{1}{\pi} \int_{\mathbb R^2} \ dq \ dp\  \exp 
 \left
 (- q^2 - p^2+ \frac{g^2}{(p^2 + \frac{2g^2}{q^2})}-
  \frac{g^2}{q^2}\right)
 \edq
 
 We shall now perform estimates of $F_{C}(t)$ and $\tilde F_{C}(t,g)$ and a fine analysis
 of the behaviour of $F_{C}(t,g)$ in the
 neighborhood of $t=0$.
 
 \begin{proposition}
(i)  $F_{C}(t,g)$ is $\pi$-periodic and we have the following estimates:
 \beq
 e^{-2g}\le F_{C}(t)\le \inf \left( 1, \sqrt 2 \exp(\frac{-g\vert \sin t \vert}
 {\sqrt{1+ \sin^2 t}})\right)
 \edq
 (ii) $\tilde F_{C}(t,g)$ is $\pi$-periodic and we have the following uniform lower bound:
 \beq
 \label{3.8}
 \tilde F_{C}(t,g)\ge 1-2g\sqrt 2
 \edq
 (iii) If  $g\ge 1/\sqrt 2$, then $\tilde F_{C}(t,g)$ attains at $t=\pi/2$ its minimum which
  equals 0.
 \beq
 \label{3.9}
  g \ge \frac{1}{\sqrt 2}\ \Longrightarrow \ \tilde F_{C}(\frac{\pi}{2},g)=0
 \edq
 \end{proposition}
 
 The Proof of Proposition 4.3 is postponed to the Appendix.

 \includegraphics{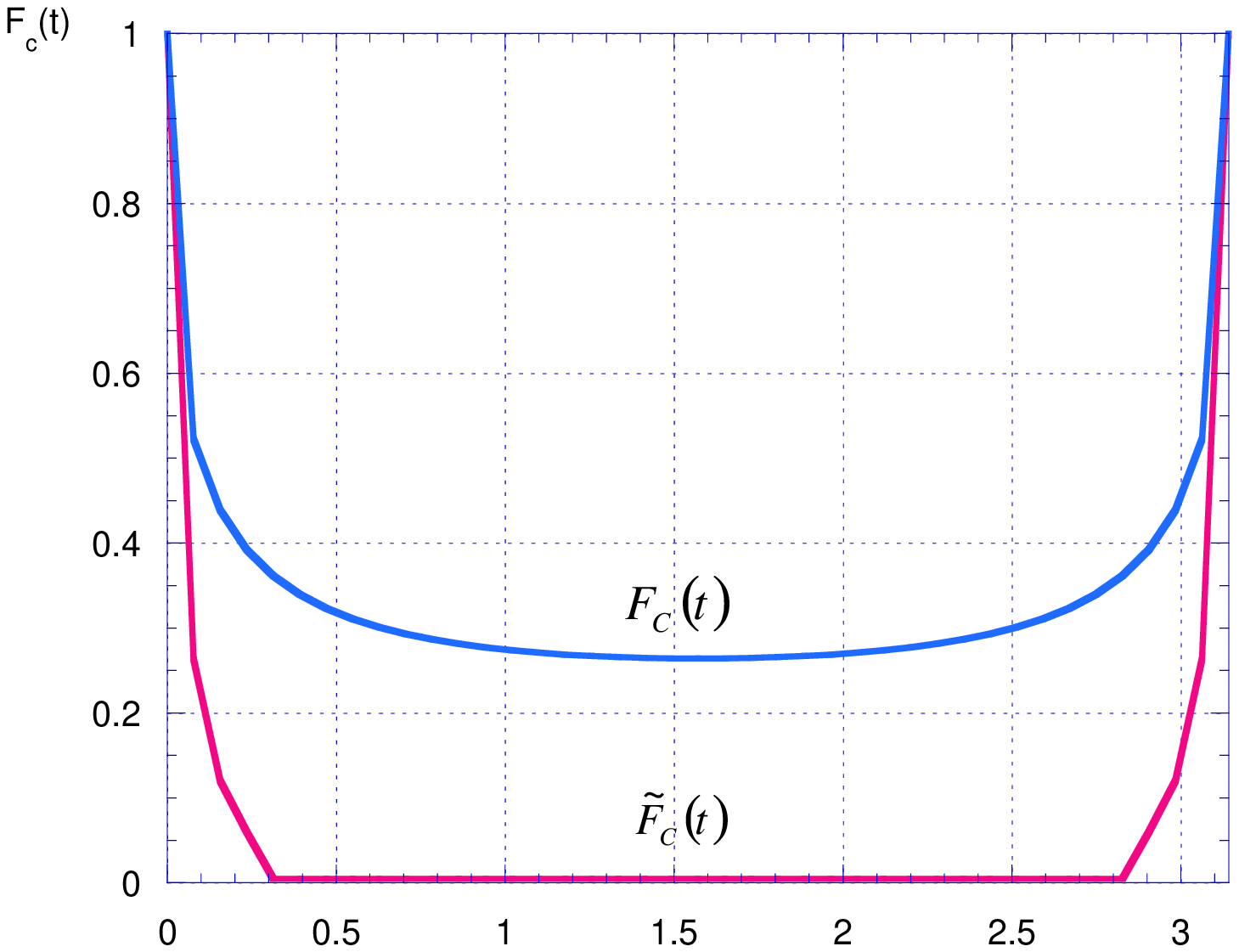}
 \begin{center}
 Fig. 2 : Classical Fidelity (Stable Case, g=1)
 \end{center}

 \begin{remark}
 The lower bound in Proposition 4.3 is uniform in $t$. It shows, as expected, that the smaller
 $g$ is, the closer to 1 is the Classical Fidelity. The upper bound shows that when g is large,
 then the Classical Fidelity can attain very small values in the interval between the recurrence 
 values $t= k\pi, \qquad k \in \mathbb Z$. Thus although being $\pi$-periodic in time,
 and thus {\bf strongly recurrent}, the Classical Fidelity can become very small for some
 values of t when g is large enough.
 \end{remark}

 We now consider the ``generic'' Classical Fidelity, independent of $g$ which appears naturally
 under the scaling $p \to (g\sqrt 2)^{1/2}p, \quad q \to (g\sqrt 2)^{1/2}q$. Then we 
 consider a ``rescaled'' distribution function:
 \beq
 \label{3.4}
 \rho(p,q) = G(p,q)\equiv \frac{1}{(\pi g \sqrt 2)^{1/2}} \exp \left( - 
 \frac{p^2 +q^2}{2g\sqrt 2}\right)
\edq
Under this scaling, we get a $g$-independent fidelity which is actually
\beq
F_{C}(t)\equiv F_{C}(t, \frac{1}{\sqrt 2})
\edq
We now give the precise behavior of $F_{C}(t)$ as $t \simeq 0$:

\begin{proposition}
Let $A>0$ any positive constant. Then, as $t\to 0$, we have:
$$F_{C}(t)\sim e^{-A\vert t \vert}$$
\end{proposition}

The Proof of Proposition 4.5 is postponed to the Appendix.

\begin{remark}
This estimate of $F_{C}(t)$ in the neighborhood of $t=0$ shows a sharp decay, faster than 
exponential from the initial value $F_{C}(0)=1$ (and due to the $\pi$-periodicity in time,
this will reproduce at $t=k\pi, \quad \forall k \in \mathbb Z$). Thus the Classical Fidelity
has cusps at those values of t.
 Proposition (4.5) (ii) shows that for $g$ small enough, then the Classical Stability function
 is very close to 1, as expected. Proposition (4.5) (iii) shows that for g large enough, then
 this Classical Fidelity function vanishes in the interval $t \in [0, \pi]$, which expresses 
 that at this point the Classical Fidelity is very bad.
 \end{remark}
 
 \bigskip
 Let us now present the curves $F_{C}(t)$ and $\tilde F_{C}(t)$ on the same diagram,
  for $t \in [0, \pi]$; (these curves are obviously $\pi$-periodic).
  
  \bigskip
 \includegraphics[width=15cm]{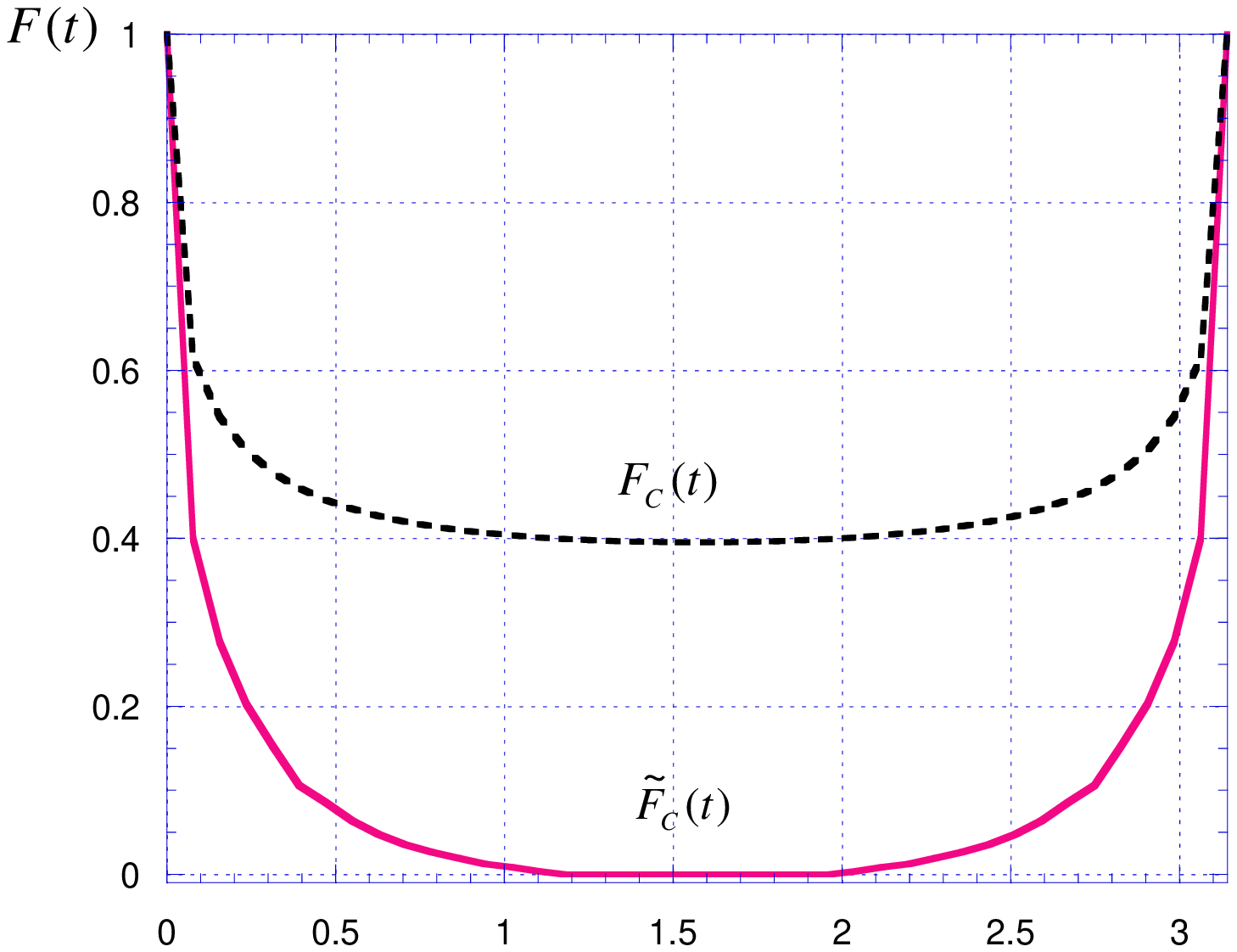}
 \begin{center}
 Fig. 3 : Classical Fidelity Functions (Stable Case)
 \end{center}

 \mysection{The Quantum Fidelity for $\epsilon =-1$}
 The most general form of solution $z_{-1}(t)$ of 
 $\ddot z -z =0$
 is 
 \beq
 \label{4.2}
 z_{-1}(t)= (a+ib)\cosh t + (c+id)\sinh t
 \edq
 $a,\ b,\ c,\ d$ being real constants. Again we assume that the Wronskian of $z$ and $\bar
  z$
 equals $2i$, which yields (\ref{2.2}).
 \\
 Let $z_{-1} := e^{u_{-1}+i\theta_{-1}}$ be the polar decomposition of $z_{-1}$.
  Thus $u_{-1}(t)$ and $\theta_{-1}(t)$ are 
 {\bf real} functions of $t$, and satisfy:
 \beq
 \label{4.3}
 \dot \theta_{-1}(t)= e^{-2u_{-1}(t)}
 \edq
 Here $\theta_{-1}(t)$ is such that
 \beq
 \label{4.4}
 \tan \theta_{-1}(t) = \frac{b \cosh t+ d \sinh t}{a \cosh t+c \sinh t}
 \edq
 \\
 
 Here the important fact is that the quantum propagators for the Inverted Harmonic (perturbed
 or unperturbed) Oscillator are given in terms of the Quantum Propagators for the 
 NON-INVERTED
 Harmonic Oscillator (perturbed or unperturbed). The time functions that enter this
 decomposition are precisely the functions $u_{-1}, \quad \theta_{-1}$ defined above
  (formula (\ref{4.4})). 
 It was established in Section 2, together with the resulting Quantum Fidelity for a 
 conveniently chosen set of reference states $\psi$. We have:
 
 \beq
 \label{4.5}
G_{Q}(t,g)\equiv \vert F_{Q}^{-1}(t,g)\vert=\vert \langle \psi_{0},
 e^{i\tilde\theta_{-1}(t) H_{0}^+}\psi_{0}  \rangle \vert
 \edq
 Thus the time dependance of $F_{Q}^{-1}(t,g)$ is governed by the 
 function $\tilde\theta_{-1}(t)\equiv \theta_{-1}(t)-\theta_{-1}(0)$ which is independent
 of $g$.
 
  \bigskip
 {\bf Study of $\tilde \theta_{-1}(t)$}\\
 
 Let us choose $b=0$ for simplicity, so that $\theta_{-1}(0)=0$ and thus $\tilde\theta_{-1}
 (t)=\theta_{-1}(t)$ in this case. Then 
 \beq
 \tan \theta_{-1}(t) = \frac{\sinh t}{a(a \cosh t+c \sinh t)}
 \edq
 which converges exponentially fast to $\frac{\pm 1}{a(a\pm c)}$ as $t \to \pm \infty$.\\
 This implies that the Quantum Fidelity never recurs to 1 in this case but tends exponentially
 fast to some constants at $\pm \infty$.
 \beq
 \cos (2 \theta_{-1}(t)) \to \frac{a^2(a\pm c)^2 -1}{a^2(a\pm c)^2 +1}
 \edq
 as $t \to \pm \infty$, exponentially fast, so that in the case $g=1$, we get a very simple 
 form of $G_{Q}(t,1)$. Let us draw the function
  $t \mapsto G_{Q}(t,1)$, in the particular case $a=-c=1$:
  
  \bigskip 
  \includegraphics{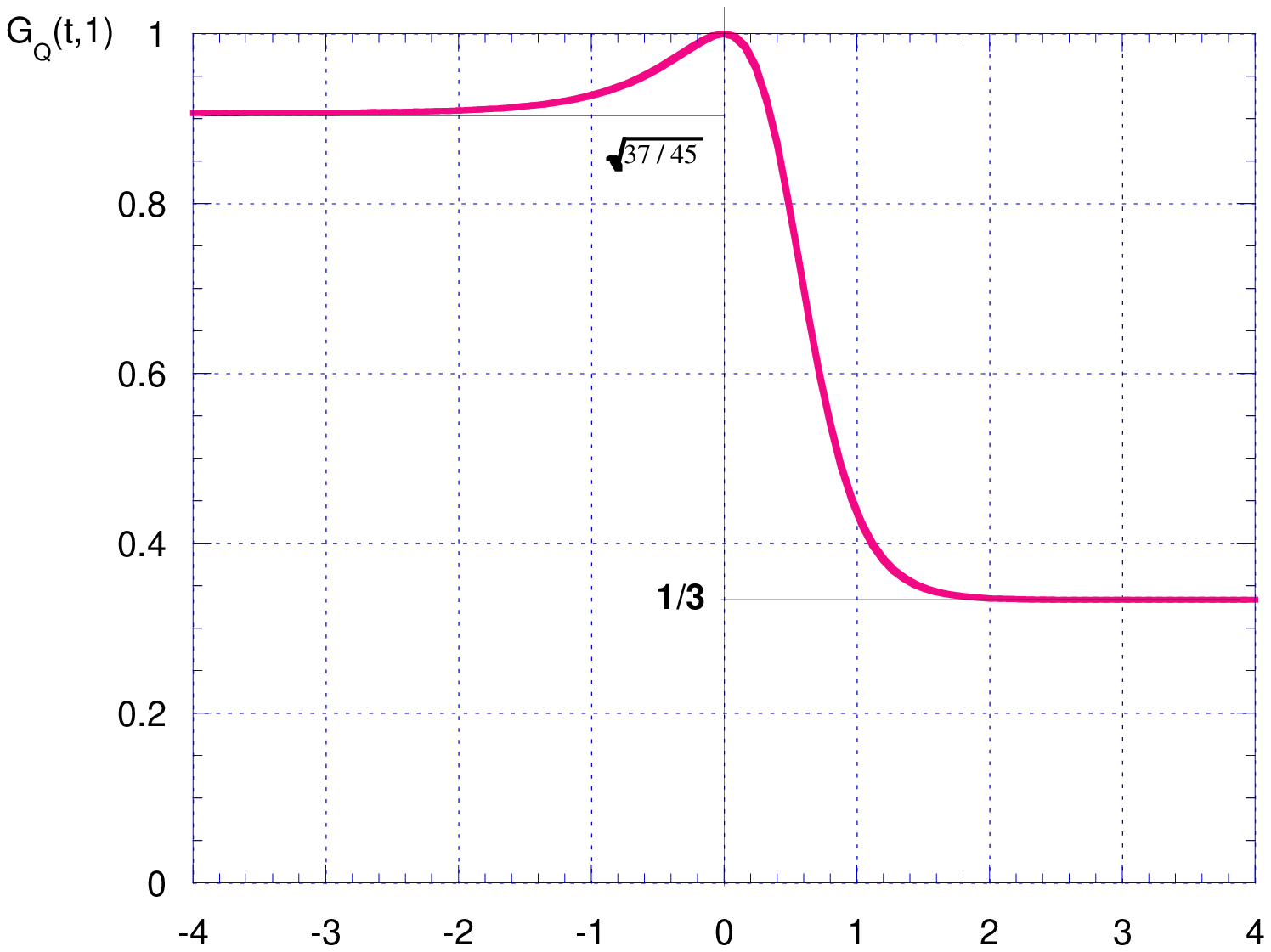}
  \begin{center}
  Fig. 4 : Quantum Fidelity (Inverted Osccilator)
  \end{center}
  
  \bigskip
  In the case $a=d=1,\  b=c=0$, we get
  \beq
  \tan \theta_{-1}(t) = \tanh t
  \edq
  so that we have a more symmetrical behavior between the future and the past of the Quantum
  Fidelity:
  \beq
  G_{Q}(t,1)^2 =  \frac{5}{9}+ \frac{4}{9 \cosh 2t}\sim \frac{5}{9} + \frac{8}{9
  }e^{-2\vert t \vert}
  \edq
   The graph of $G_{Q}(t,1)$ is then as follows:

  \bigskip
   
  \includegraphics[width=15cm]{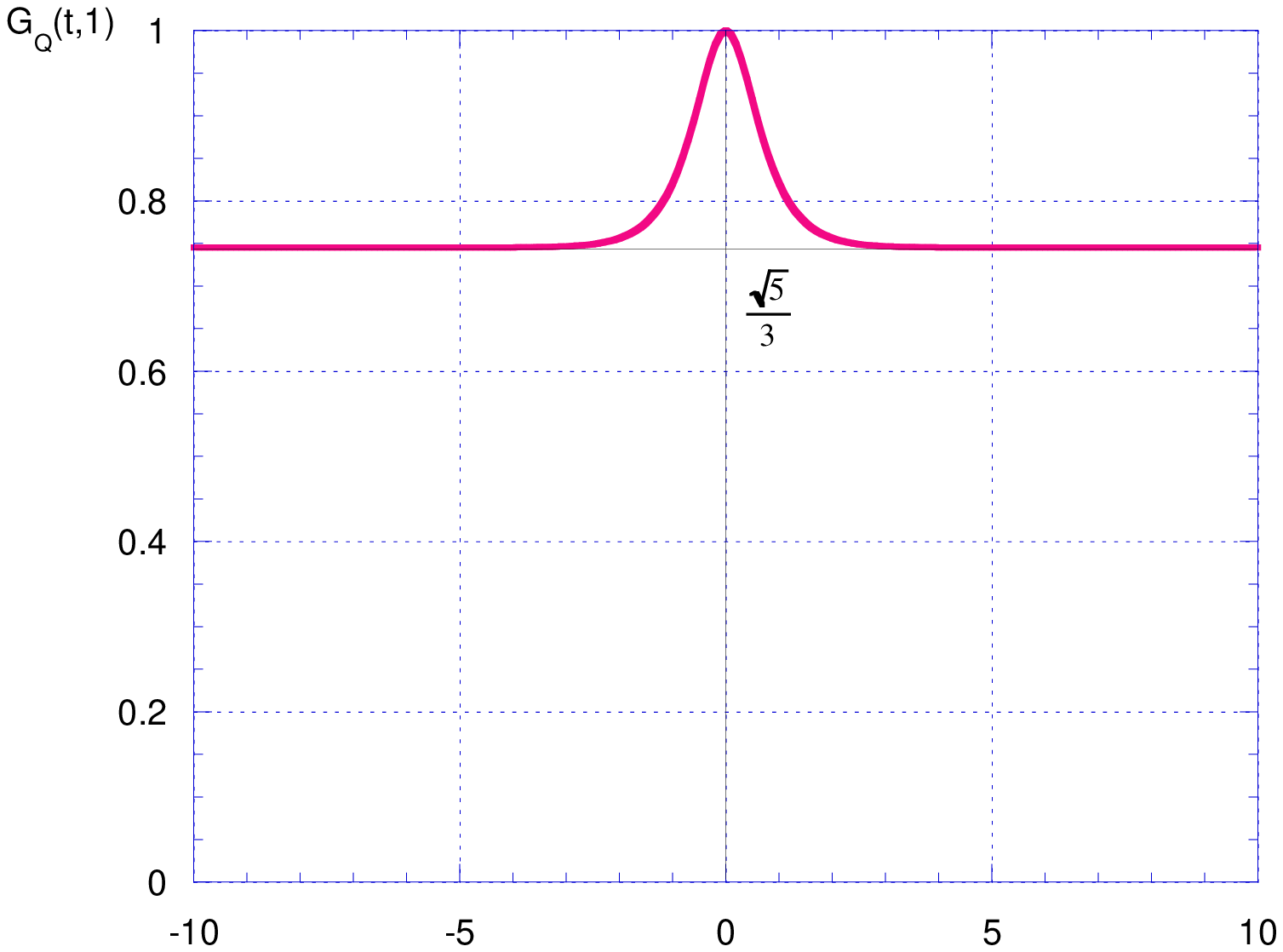}
   \begin{center}
   Fig. 5 : Quantum Fidelity (Unstable Symmetric Case)
   \end{center}
   
   \begin{remark}
   The Quantum Fidelity, which we have denoted $G_{Q}(t,1)$ in this case is always bounded
   from below by 1/3, uniformly in $t$. A similar result holds true for $G_{Q}(t, \sqrt 3)$, 
   and for $G_{Q}(t, \sqrt{10})$.
   \end{remark}
 
\section{Comparison with the Classical Fidelity for $H_{g}^{-}$}
 
Here again we have that the classical trajectories for $H_{g}^{-}$ can easily be deduced
 from that for $H_{0}^{-}$, and that the natural scaling $y(t)= (g \sqrt 2)^{1/2}y(t)$
 makes them independent from $g$.
 
 \begin{proposition}
 Let $z_{-1}(t)$ be given by (\ref{4.2})
 with Wronskian of $(z_{-1}, \ \bar z_{-1})$ being equal to $2i$ which yields (\ref{2.2}).
  If $z_{-1}
 = e^{u_{-1}+i\theta_{-1}}$ is the polar
  decomposition of $z$, with $\theta_{-1}(0)=0$, then , given initial data $q,\ p$ with
   $q \ne 0$,
  \beq
  \label{5.1}
  z_{-1}(t)= q \cosh t + p \sinh t + i \frac{\sinh t}{q}
  \edq
  is such that $\Re z_{-1}(t):= x(t)$ obeys the differential equation $\ddot x - x =0$, and
  \beq
  x(0)=q, \ \dot x(t)= p
  \edq
  and 
  \beq 
  \label{5.2}
  y(t):= \vert z(t)\vert
 \edq
 is a trajectory for $\frac{p^2}{2} - \frac{q^2}{2}+ \frac{1}{q^2}$, namely obeys
 \beq
 \label{5.3}
 \ddot y -y -\frac{1}{y^{3}}=0
 \edq
 with the same initial data as $x(t)$:
 \beq
 y(0)=q,\ \dot y(0) = p
 \edq
 \end{proposition} 
 
 \begin{remark}
 We recover the general trajectory for $H_{g}^{-}$ by simply rescaling the initial data $(q,\ 
  p)$ by a factor $(g \sqrt 2)^{-1/2}$, which provides solutions of
  \beq
  \label{5.4}
  \ddot y'-y' -\frac{2g^2}{y'^3}=0
  \edq
  of the form:
  \beq
  y'(t)= \left((q\cosh t + p \sinh t)^2+\frac{2g^2 \sinh^2 t}{q^2}\right)^{1/2}
  \edq
 \end{remark}
 
 Proof of Proposition 6.1:
 \\
 It is essentially equivalent to that of Proposition 4.1.
 
 \bigskip 
 We now define the Classical Fidelity in terms of distribution functions $\rho(p,q)$ as in 
 Section 2 for the stable case. We denote them by $G_{C}(t,g)$ and $\tilde G_{C}(t,g)$ 
 for $\rho = G(p,q), \ X(p,q)$ respectively, (namely the Gaussian and characteristic functions
 of  phase-space variables $(q,p)$ that we have introduced in Section 4). However here,
 we assume that
 \beq
 \label{5.5}
 X(p,q):= \frac{1}{\sqrt{3\pi}} \ \chi(p^2+q^2\le 3\pi)
 \edq
 
 \begin{proposition}
 We have the following uniform in $t$ lower bounds for $G_{C}(t,g),\ \tilde G_{C}(t,g)$:
 \beq
 \label{5.6}
 G_{C}(t,g)\ge e^{-2g}
 \edq
 \beq
 \label{5.7}
 \tilde G_{C}(t,g)\ge 1-\frac{2g\sqrt 2}{3}
 \edq
 \end{proposition}
 
 The Proof is postponed to the Appendix.

 \begin{remark}
 Again for $g$ small the Classical Fidelities $G_{C}(t,g),\ \tilde G_{C}(t,g)$ remain 
 close to 1 uniformly in time. We now explain why we have chosen the characteristic function
 of the ball $\left\{ (p,q)\in \mathbb R^2: \ p^2+q^2\le 3\right\}$ instead of the ball
  of radius 1.
 We have generic
 {\bf g-independent} Classical Fidelities that we denote $G_{C}(t),\ \tilde G_{C}(t)$ by
 using the rescaled variables $p(g\sqrt 2)^{1/2}, \ q(g\sqrt 2)^{1/2}$. Under the rescaling
 of the Gaussian functions and characteristic functions used in the definitions, we find simple
 estimates as seen below. 
 \end{remark}
 
 \begin{proposition}
 Let $x(t),\ y(t)$ be as in Proposition 6.1. Then define:
 \beq
 \label{5.30}
 G_{C}(t):= \frac{1}{\pi}\ \int \ dq\ dp \ \exp\left(-\frac{x^2(t)+\dot x(t)^2
 +y^2(t)+\dot y(t)^2}{2}\right)
 \edq
 \beq
 \label{5.31}
 \tilde G_{C}(t):= \frac{1}{3\pi}\ \int\ dp\ dq\ \chi(x^2(t)+\dot x(t)^2\le 3)
 \ \chi(y(t)^2+\dot y(t)^2\le 3)
 \edq
(i)  We have the uniform lower bounds:
 \beq
 G_{C}(t)\ge e^{-\sqrt 2}
 \edq
 \beq
 \tilde G_{C}(t)\ge \frac{1}{3}
 \edq
 (ii) As $t \to \infty$, the limiting value of $\tilde G_{C}(t)$ satisfies the following estimate:
 \beq
 \label{5.20}
 \tilde G_{C}(\infty)\le \frac{2\arctan (\sqrt{17})}{\pi}-
 \frac{\sqrt{17}}{9\pi}
 \simeq 0.497
 \edq
 \end{proposition}
 
 The Proof is postponed to the Appendix.
 \\
 
 This shows that the Classical Fidelity in this case doesn't decay to zero but instead stays 
 bounded below by some constant.
 We now present on the  graphic below the graph of $G_{C}(t)$:

 \includegraphics{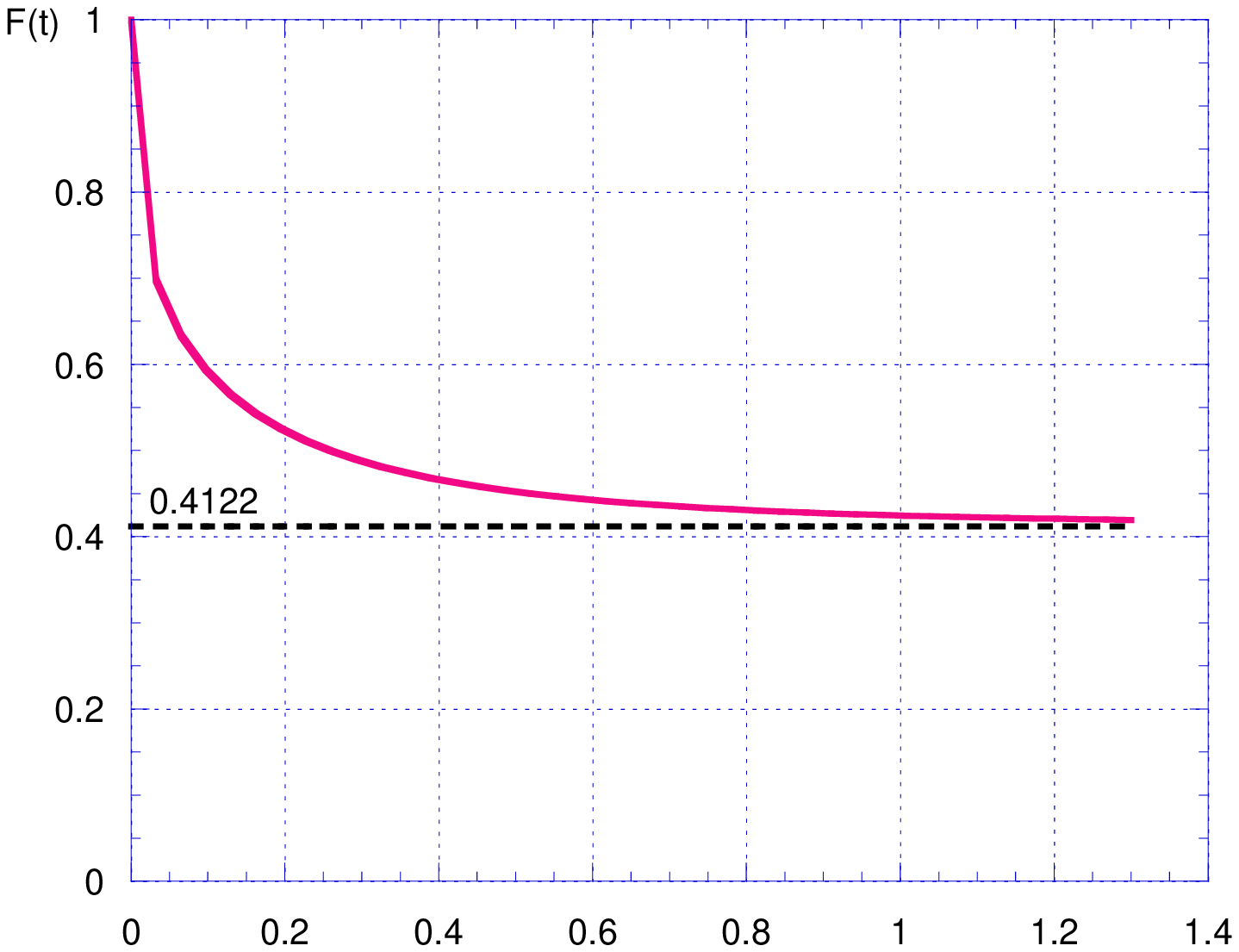}
 \begin{center}
 Fig. 6 : Classical Fidelity (Unstable Case)
 \end{center}

 Owing to equations (\ref{5.30}) and (\ref{5.31}), the Classical Fidelities have asymptotic
 values at $t \to \infty$ (thus $\tau := (\cosh 2t)^{-1}=0$) equal respectively to
 \beq
 G_{C}(\infty)= \frac{1}{\pi}\ \int_{\mathbb R^2}\ du\ dv\ \exp\left(-u^2
 -v^2-\frac{1}{2u^2}+\frac{1}{(u+v)^2+\frac{1}{u^2}}\right)\simeq 0.414
 \edq
 \beq
 \tilde G_{C}(\infty)=\frac{1}{3\pi}\int_{\mathbb R^2}\ du\ dv\ \chi(u^2+v^2
 \le 3)\ \chi\left(u^2+v^2+\frac{1}{u^2}-\frac{2}{(u+v)^2+ \frac{1}{u^2}}\le
 3 \right)\simeq 0.497
 \edq
 
 \bigskip
 Furthermore let us give simple expressions of the $g$-dependent Classical Fidelities, using
 the scaling noted above and the $g$-independent functions $y(t)$ given in Proposition 6.1.
 It appears that these expressions are more suitable for the numerical investigation of the
 dependence in parameter $g$ of $G_{C}(t,g)$ and $\tilde G_{C}(t,g)$:
 
 \beq
 \label{5.32}
  \frac{g \sqrt 2}{\pi}\ \int_{\mathbb R^2}\ du\ dv \exp\left(-g\sqrt 2
 \left[u^2+v^2+\frac{1}{u^2}-\frac{2}{u^2(1+\tau)+v^2(1-\tau)+(1-\tau)\frac{1}
 {u^2}+2uv \sqrt{1-\tau^2}}\right]\right)
 \edq
 $$=G_{C}(t,g)$$
 \beq
 \label{5.33}
\tilde G_{C}(t,g)= \frac{g\sqrt 2}{3\pi}\ \int_{\mathbb R^2}\ du\ dv\ \chi(u^2
 +v^2\le \frac{3}{g\sqrt 2})\ 
 \edq
 $$\times\chi\left(u^2+v^2+\frac{1}{2u^2}-\frac{1}{
 u^2(1+\tau)+v^2(1-\tau)+(1-\tau)\frac{1}{u^2}+2uv \sqrt{1-\tau^2}}\le \frac{3}
 {g\sqrt 2}\right)$$
 
 Recall that $$\tau := \frac{1}{\cosh 2t}$$
 
 We now draw on the same picture the Classical Fidelity functions $G_{C}(t)$ and $\tilde
 G_{C}(t)$: (we have not been able to prove rigorously that they actually monotonically 
 decay to these asymptotic value as $t$ goes to $\infty$). Next we draw on the same graphic
  the Classical and Quantal Fidelities for the specific value $g=1$.
 
% \bigskip
 
 \includegraphics{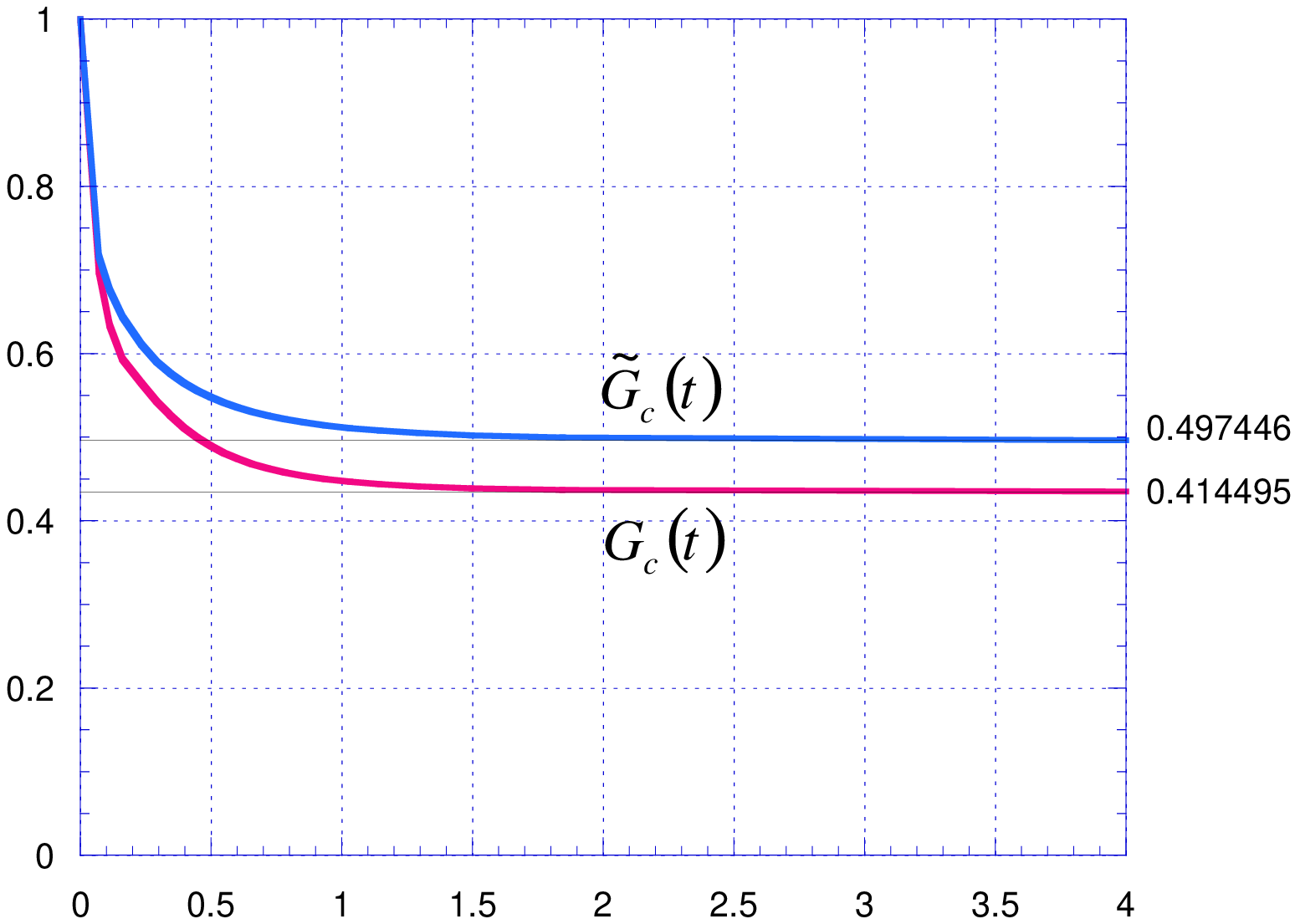}
 \begin{center}
 Fig. 7 : Classical Fidelity (Unstable Case g=1)
 \end{center}
 
\bigskip
\includegraphics{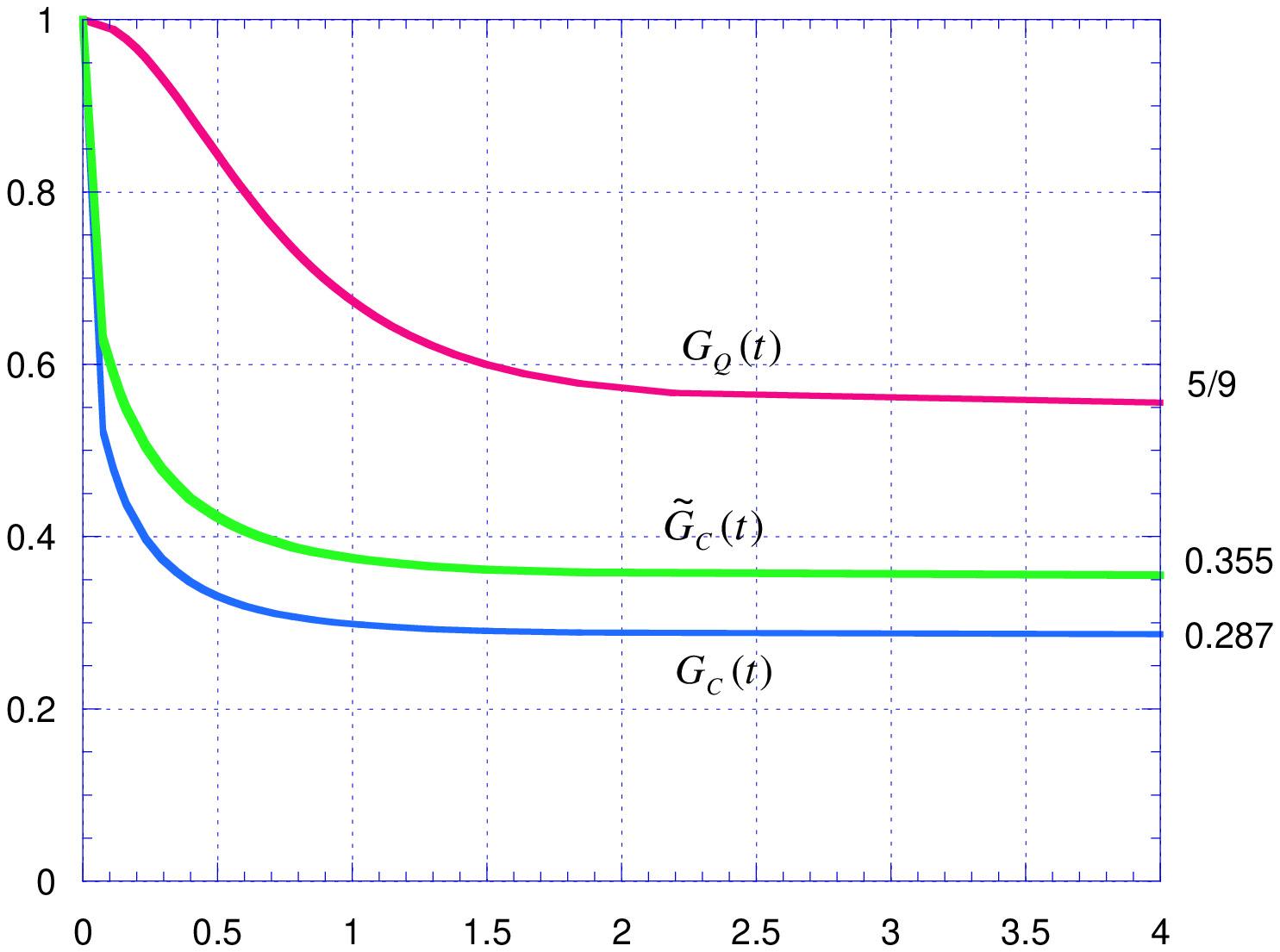}
\begin{center}
Fig. 8 : Quantum Classical fidelity Comparison (Unstable case g=1 )
\end{center}
 
 \section{CONCLUDING REMARKS}
 
 In this paper we show the behaviour in time of Quantum as well as Classical Fidelities for a very
 special class of Hamiltonians for which the quantum dynamics is exactly solvable in terms
 of the classical one. For $\epsilon = +1$, the motion is stable and manifests strong periodic
 recurrences in the classical as well of quantum evolution. It has been exhibited as well in
 time-periodic systems in \cite{co1}, and recurrences were already shown for another class 
 of systems in \cite{sala}.\\

 For $\epsilon =-1$, the classical motion is unstable, although not chaotic. In this case a
 decrease in time (at $\pm \infty$) occurs for the Quantum as well as the Classical Fidelity 
 functions. However they do not decay to zero, but instead both remain bounded from below.
 This is important to have here an explicit example where the fidelities (classical as well as
 quantum) do not decay to zero. This is in contrast with the general ``chaotic'' situation 
 where the fidelities are generally claimed (although not proven) to decay rapidly to 0.\\
   
 Qualitatively the behaviour of the Quantum and Classical Fidelities show strong resemblance, 
 except in the neighborhood of $t=0$ (and of $t= k\pi,\quad k \in \mathbb Z$ for the
 stable case because of periodicity). 
 Namely, instead of being smooth at $t=0$ the classical fidelities display a cusp.
 We believe that taking better classical distribution
 functions in phase-space, and restoring the small parameter $\hbar$ should correct this 
 defect. One might think taking the Wigner functions of the reference wavepackets; however,
 it is known that only Gaussian wavepackets provide  nonnegative Wigner functions which
 thus
 mimic a ``probability distribution in phase space'', and this is not the case here.
 \\
 
 Finally we note that, at least in the cases $g=1,\ \sqrt 3, \ \sqrt{10}$ where we have
 explicit behaviours of the Quantum Fidelities, these functions remain above the Classical Fidelity
 functions, in the stable as well as unstable case. However for a better understanding of the
 correspondence quantum/classical, one has to restore the parameter $\hbar$ and use the
 semiclassical approach. This is first considered in a rigorous framework (and in a much more
 general setting) in \cite{coro}.

 \section{APPENDIX}
 {\bf Proof of Proposition 4.3}\\
(i)  Let us start with the lower bound. Clearly $y'(t)^2+ \dot y'(t)^2 \le q^2+p^2+
 \frac{g^2}{q^2}$, so that 
 \beq
 F_{C}(t)\ge \frac{1}{\pi} \int_{\mathbb R^2} \ dq\ dp\ e^{-p^2-q^2- 
 \frac{g^2}{q^2}}= e^{-2g}
 \edq
 uniformly in $t$, where we use the explicit formula:
 \beq
\label{3.6}
\int_{\mathbb R} \ dx\ \exp\left(-Ax^2-\frac{B}{x^2}\right)= \sqrt{\frac{\pi}
{A}}e^{-2\sqrt{AB}}
\edq
 
  Of course equ. (\ref{3.3}) is a much better lower bound, but we have 
 unfortunately no exact computation of the integral. Let us now consider the upper bound.\\
 
 We denote $q= r\cos \alpha$, $p=r \sin \alpha$, and assume that $r \ne 0$. Then
 \beq
 y'(t)^2+ \dot y'(t)^2 = r^2+ \frac{2g^2}{r^2 \cos^2 \alpha}- \frac{2g^2}{r^2 \cos^2
 (t-\alpha)+ \frac{2g^2}{r^2\cos^2 \alpha}\sin^2 t}
 \edq
 
$$ \ge r^2 \cos^2 (t-\alpha) + \frac{2g^2 \sin^2 t}{r^2 \cos^2 \alpha}$$
This implies:
\beq
q(t)^2+p(t)^2+y'(t)^2+ \dot y'(t)^2 \ge p^2+q^2+ (q\cos t + p \sin t)^2 + \frac{2g^2
\sin^2 t}{q^2}
\edq
Whence
\beq
F_{C}(t)\le \frac{1}{\pi} \int \ dq \ dp \ \exp \left(- \frac{p^2+q^2}{2}
-\frac{1}{2}(q\cos t + p \sin t)^2- \frac{ g^2\sin^2 t}{q^2}\right)
\edq
The integral over $p$ can be performed easily:
\beq
\int dp \ \exp \left(-\frac{1}{2}(1+\sin^2 t)\left [p+\frac{q\cos t \sin t}
{1+\sin^2 t}
\right]^2\right)= \sqrt{\frac{2\pi}{1+\sin^2 t}}
\edq
Thus we are left with an integral over $q$ of the form:
\beq
\int dq\ \exp\left(\frac{-q^2}{1+\sin^2 t}-\frac{ g^2\sin^2 t}{q^2}\right)=
\sqrt{\pi (1+\sin^2 t)}\exp\left( \frac{-g\vert \sin t \vert}{\sqrt{1+\sin^2 t}}
\right)
\edq
This yields the final upper bound which is of course not optimal in the neighborhood of
$t=0 \quad (\mbox{mod }\ \pi)$.\\

 (ii) \beq
\tilde F_{C}(t,g)= \frac{1}{\pi}\int_{\mathbb R^2}\ dq\ dp\ \chi(p^2+q^2\le 1)\ 
 \chi(\dot y'(t)^2+y'(t)^2 \le 1)
 \edq
$$ = \frac{1}{\pi}\int \ dq \ dp\ \chi(q^2+p^2\le 1)\ \chi(p^2+q^2 +\frac{2g^2}
 {q^2}-\frac{2g^2}{y'^2(t)}\le 1)
 \ge \frac{1}{\pi}\int \ dq\ dp \ \chi(p^2+q^2+\frac{2g^2}{q^2}\le 1)$$
 $$=\frac{1}{\pi}\int \ dq \ dp \ \chi(p^2+ (q-\frac{g \sqrt 2}{q})^2
 \le 1-2g \sqrt 2)$$
 $$=2 \frac{(g \sqrt 2)^{1/2}}{\pi}\int_{-\infty}^{+\infty}\ dp 
 \int_{0}^{+\infty}\ dq' \ \chi(p^2+g\sqrt2(q'-\frac{1}{q'})^2\le
 1-2g\sqrt 2)$$
 $$= \frac{(g\sqrt 2)^{1/2}}{\pi}\int_{0}^{+\infty}\ dq'\int_{-\infty}^{+\infty}\  dp\ 
 (1+\frac{1}{q'^2})\ \chi(p^2 +g\sqrt 2(q'-\frac{1}{q'})^2\le 1-2g \sqrt 2)$$
 $$= \frac{g\sqrt 2}{\pi}\int_{\mathbb R^2}\ dp' \ du \ \chi(p'^2+u^2\le
 \frac{1-2g\sqrt 2}{g\sqrt 2})= 1-2g\sqrt 2$$
 where we have used the changes of variables $q = (g\sqrt 2)^{1/2}q',\quad  
 u:= q'-\frac{1}{q'}$
 \\
 
 (iii) Consider the rescaled variables $q' :=(g\sqrt 2)^{-1/2}q, \quad p' := (g\sqrt 2)^{-1/2}
 p$; then
 \beq
 \tilde F_{C}(t,g)= \frac{g\sqrt 2}{\pi}\int_{\mathbb R^2}\ dq'\ dp'\ 
 \chi(p'^2+q'^2\le \frac{1}{g\sqrt 2})\ \chi(q'^2+p'^2+\frac{1}{q'^2}-\frac{1}
 {y'^2(t)}\le \frac{1}{g\sqrt 2})
 \edq
 But 
 \beq
 \chi(y'^2(t)+\dot y'^2(t)\le \frac{1}{g\sqrt 2})\le \chi(p'^2+\frac{1}{q'^2}\le
 \frac{1}{g\sqrt 2})
 \edq
 so that 
 \beq
 \tilde F_{C}(\frac{\pi}{2}, g)\le \frac{g\sqrt 2}{\pi}\int_{\mathbb R^2}\ 
 dp'\ dq' \ \chi(p'^2+q'^2\le \frac{1}{g\sqrt 2})\ \chi(p'^2+ \frac{1}{q'^2}\le
 \frac{1}{g \sqrt 2})
 \edq
 Now, for $g\ge \frac{1}{\sqrt 2}$ the domains
 $$\left\{ (q',p')\in \mathbb R^2:\ p'^2+q'^2\le \frac{1}{g\sqrt 2}\right\},\ 
 \mbox{and}\ 
 \left\{(q',p')\in \mathbb R^2: \ p'^2+\frac{1}{q'^2}\le \frac{1}{g\sqrt 2}
 \right\}$$
 have no  common domain of positive area, so that 
 \beq
 \tilde F_{C}(\frac{\pi}{2}, g)=0
 \edq
 \sq
 \\

{\bf Proof of Proposition 4.5}\\
Denote: 
\beq
Q(p,q):= p^2+q^2+ y^2(t)+\dot y^2(t)= 2(p^2+q^2)+\frac{1}{q^2}-\frac{1}{y^2(t)}
\edq
Let us expand $Q(p,q)$ near $t=0$ up to order 2. We get:
\beq
Q(p,q)\simeq 2(p^2+q^2)+2t \frac{p}{q^3}-\frac{t^2}{2q^2}+ \frac{p^2 t^2}{q^4}
+\frac{t^2}{q^6}
\edq
Therefore
\beq
e^{-\frac{Q(p,q)}{2}}\lesssim e^{-(p+\frac{t}{2q^3})^2 -q^2 +\frac{t^2}{4q^2}
-\frac{t^2}{4q^6}}
\edq
We thus have:
\beq
\label{3.5}
\frac{1}{\pi}\int \ dp\ dq\ e^{-Q(p,q)/2}\lesssim \frac{1}{\sqrt \pi}\int \ dq
\ \exp\left(-q^2 +\frac{t^2}{4q^2}-\frac{t^2}{4q^6}\right)
\edq
Let us divide the integration domain into two parts:\\
$\bullet\ q^4 \le \frac{1}{1+A^2}$\\
$\bullet \ q^4 \ge \frac{1}{1+A^2}$\\

$A>0$ being an arbitrary constant. Then in the first domain, we have:
\beq
\frac{1}{4q^2}-\frac{1}{q^6}\le -\frac{A^2}{4q^2}
\edq
so that the corresponding contribution to equ. (\ref{3.5}) is approximated by:
\beq
\frac{1}{\sqrt \pi}\int_{q^4 \le \frac{1}{1+A^2}}\ dq \ e^{-q^2-\frac{t^2 A^2}
{4q^2}}
\edq
In the second domain we have:
\beq
-q^2+\frac{t^2}{4q^2}\le -q^2\left (1-\frac{(1+A^2)^2 t^2}{4}\right)-\frac{A^2 t^2}
{4q^2}
\edq
so that the corresponding contribution to equ. (\ref{3.5}) is 
\beq
\lesssim \frac{1}{\sqrt \pi}\int_{q^4\ge \frac{1}{1+A^2}} \ dq \ \exp\left(-q^2 (1
-\frac{t^2 (1+A^2)^2}{4})-\frac{A^2 t^2}{4q^2}\right)
\edq
Summing up the two contributions we get as $t\simeq 0$
\beq
\frac{1}{\pi} \int \ dp\ dq\ e^{-\frac{Q(p,q)}{2}}\lesssim \frac{1}{\sqrt \pi}
\int dq \exp \left(-q^2(1-\frac{t^2 (1+A^2)^2}{4})-\frac{t^2 A^2}{4q^2}\right)
\edq
$$\simeq\left(1-\frac{t^2 (1+A^2)^2}{4}\right)^{-1/2}\exp \left(-A\vert t \vert
(1-\frac{t^2 (1+A^2)^2}{4})^{1/2}\right)
$$
using the explicit formula (\ref{3.6}) for any positive constants $A,\ B$. Now taking the 
dominant behavior of the RHS of 
equ.(75)
as $t \simeq 0$ yields the result.
\sq
\\

{\bf  Proof of Proposition 6.3}\\
 Due to conservation of energy for Hamiltonian $H_{g}^-$, we have:
 \beq
 \label{5.8}
 \dot y'^2(t)-y'^2(t)+\frac{2g^2}{y'^2(t)}= p^2-q^2+\frac{2g^2}{q^2}
 \edq
 so that:
 \beq
 \label{5.9}
\dot y'^2(t)+y'^2(t) = 2y'^2(t)+ p^2-q^2+\frac{2g^2}{q^2}-\frac{2g^2}{y'^2(t)}
\edq
$$=\cosh 2t (q^2+p^2+\frac{2g^2}{q^2})+ 2pq\sinh 2t-\frac{2g^2}{y'^2(t)}$$
$$= \cosh 2t \left((p+q\tanh 2t)^2+\frac{2g^2}{q^2}\right) + \frac{q^2}{\cosh 2t}
-\frac{2g^2}{y'^2(t)}$$
We perform the following changes of variables:
\beq
\label{5.10}
v:= \sqrt{\cosh 2t}(p+q\tanh 2t), \quad u:= \frac{q}{\sqrt{\cosh 2t}}
\edq
This yields:
\beq
2y'^2(t)= v^2(1-\tau)+u^2(1+\tau)+\frac{2g^2}{u^2}(1-\tau)+2uv \sqrt{1-\tau^2}
\edq
and
\beq
\label{5.11}
\dot y'(t)^2+y'(t)^2 = v^2+u^2+\frac{2g^2}{u^2}-\frac{4g^2}{v^2(1-\tau)
+u^2(1+\tau)+2uv \sqrt{1-\tau^2}+(1-\tau) \frac{2g^2}{u^2}}
\edq
where we have denoted 
\beq
\label{5.12}
\tau := \frac{1}{\cosh 2t}\le 1
\edq
It clearly follows from (\ref{5.11}) that
\beq
\dot y'^2(t)+y'^2(t) \le v^2+u^2+\frac{2g^2}{u^2}
\edq
Now in terms of the same variables $u,\ v$ we have:
\beq
\label{5.14}
\dot x(t)^2+x(t)^2= v^2+u^2
\edq
When passing to the new integration variables $u,\ v \in \mathbb R$, we get:
\beq
\label{5.13}
G_{C}(t,g)\ge \frac{1}{\pi}\int_{\mathbb R^2}\ du\ dv \ \exp\left(-u^2-v^2
-\frac{g^2}{u^2}\right) = e^{-2g}
\edq 
using again the formula (\ref{3.6}).\\

 Now we come to $\tilde G_{C}(t,g)$:
\beq
\label{5.14}
\tilde G_{C}(t,g)\ge \frac{1}{3\pi}\int_{\mathbb R^2}\ du\ dv\ \chi \left(
u^2+v^2+\frac{2g^2}{u^2}\le 3\right)
\edq
$$= \frac{2}{3\pi}\int_{-\infty}^{+\infty}\ dv\ \int_{0}^{+\infty}\ du
\ \chi\left(v^2+ (u-\frac{g\sqrt 2}{u})^2 \le 3-2g\sqrt 2 \right)$$
\beq
= \frac{2(g\sqrt 2)^{1/2}}{3\pi}\int_{0}^{+\infty}\ du' \int_{-\infty}^{+\infty}
\ dv\ \chi\left(v^2+
 g\sqrt 2(u'-\frac{1}{u'})^2 \le 3-2g\sqrt 2\right)
\edq
$$=\frac{(g\sqrt 2)^{1/2}}{3\pi}\int_{0}^{+\infty}\ du' \int_{-\infty}^{+\infty}
\ dv\ (1+\frac{1}{u'^2})\chi\left(v^2+
 g\sqrt 2(u'-\frac{1}{u'})^2 \le 3-2g\sqrt 2\right)$$
 We now use the new integration variable $x:=(g\sqrt 2)^{1/2}(u'-\frac{1}{u'})$ which
 yields:
 \beq
 \frac{1}{3\pi}\int_{\mathbb R^2}\ dx\ dv\ \chi(x^2+v^2\le 3-2g\sqrt 2)=
  1-\frac{2g\sqrt 2}{3}
  \edq
  \sq\\
  
{\bf Proof of Proposition 6.5}\\
 Using the same change of variable $(p,\ q)\to (u,\ v)$ as in (\ref{5.10}), we get:
\beq
G_{C}(t)\ge \int_{\mathbb R^2}\ du\ dv\ \exp\left(-u^2-v^2-\frac{1}{2u^2}\right)
= e^{-\sqrt 2}
\edq
and
\beq
\tilde G_{C}(t)\ge \frac{1}{3\pi}\ \int_{\mathbb R^2}\ du\ dv\ \chi\left(u^2
+v^2+\frac{1}{u^2}\le 3 \right)= \frac{1}{3\pi}\ \int_{\mathbb R^2}\ du\ dv\ 
\chi\left(v^2+(u-\frac{1}{u})^2 \le 1 \right)
\edq
$$= \frac{1}{3\pi}\int_{0}^{+\infty}\ du\ \int_{-\infty}^{+\infty}\ dv\ (1+
\frac{1}{u^2})\ \chi\left(v^2+(u-\frac{1}{u})^2\le 1 \right)$$
$$= \frac{1}{3\pi}\ \int_{\mathbb R^2}\ dv\ dx \ \chi(x^2+v^2\le 1)= \frac{1}{3}
$$
\sq\\

 (ii) The quantity $\dot y(t)^2+y(t)^2$ (with the scaled variables as in Proposition 6.1)
 can be rewritten as in (\ref{5.11}) using the variables $u,\ v,\ \tau$ defined by (\ref{5.10})
 and (\ref{5.12}):
 \beq
 \dot y(t)^2+y(t)^2= u^2+v^2+\frac{1}{u^2}-\frac{2}{v^2(1-\tau)+u^2(1+\tau)+
 (1-\tau)\frac{1}{u^2}+2uv\sqrt{1-\tau^2}}
 \edq
 so that as $t \to \infty$, $\tau \to 0$ very rapidly and we get
 \beq
 \tilde G_{C}(\infty)= \frac{1}{3\pi}\int_{\mathbb R^2}\ du\ dv\ \chi(u^2+v^2
 \le 3)\ \chi\left(u^2+v^2+\frac{1}{u^2}-\frac{2}{(u+v)^2+ \frac{1}{u^2}}\le
 3 \right)
 \edq
 But since 
 \beq
 \left\{ (u,\ v):\  u^2+v^2+\frac{1}{u^2}-\frac{2}{(u+v)^2+\frac{1}{u^2}}\right\}
 \subset \left\{(u,\ v):\ u^2+v^2+\frac{1}{u^2}-2u^2 \le 3 \right\}
 \edq
 \beq
 \tilde G_{C}(\infty)\le \frac{1}{3\pi}\ \int_{\mathbb R^2}\ du\ dv\ \chi(u^2
 +v^2\le 3)\ \chi\left(v^2-u^2+\frac{1}{u^2}\le 3 \right)
 \edq
 $$\le \frac{1}{3\pi}\ \int_{\mathbb R^2}\ du\ dv\ \chi(u^2+v^2\le 3)\ \chi
 \left(\frac{1}{u^2}\le 6 \right)$$
 This means that the estimate equals the part of the disk in $u,\ v$ space of radius $\sqrt 3$
 outside of the interval $\vert q \vert \le \frac{1}{\sqrt 6}$, divided by $3\pi$. This
 equals precisely
 \beq
 \frac{6\arctan (\sqrt{17})-\sqrt{17}/3}{3\pi}= \frac{2\arctan (\sqrt{17})}{\pi}
 -\frac{\sqrt{17}}{9\pi}
 \edq
 \sq\\

{\bf Acknowledgments} We thank the anonymous referee for providing useful comments
and criticisms of a first draft of this paper.

 \bibliographystyle{amsalpha}

\end{document}